\documentclass{aa}

\newcommand{\mdens}{{\rm g~cm^{-3}}}
\newcommand{\bdens}{{\rm fm^{-3}}}
\newcommand{\msun}{{\rm M}_\odot}

\usepackage{graphicx}
\usepackage{amssymb,amsmath}
\usepackage[comma]{natbib}
\usepackage{color}
\usepackage{bm}
\begin{document}

\title{Hyperons in neutron-star cores and $2\msun$ pulsar}
\author{I. Bednarek\inst{2} \and P. Haensel\inst{1} \and
J.L. Zdunik\inst{1}\and M. Bejger\inst{1}
\and R. Ma{\'n}ka\inst{2}\thanks{retired}}
\institute{N. Copernicus
Astronomical Center, Polish Academy of Sciences, Bartycka 18,
PL-00-716 Warszawa, Poland
\and
Department of Astrophysics and Cosmology, Institute of Physics,
University of Silesia, Uniwersytecka 4, 40-007 Katowice, Poland
\\{\tt ilona.bednarek@us.edu.pl,bejger@camk.edu.pl,
haensel@camk.edu.pl, jlz@camk.edu.pl}}
\offprints{I. Bednarek}
\date{Received xxx Accepted xxx}

\abstract{Recent measurement of mass of PSR J1614-2230 rules out most
of existing models of equation of state (EOS) of dense matter with high-density
softening due to hyperonization or a phase transition to quark matter
or a boson condensate. }
{We look for a solution of an apparent contradiction between the consequences
stemming from up-to-date hypernuclear data, indicating appearance of hyperons at
$\sim 3$ nuclear densities and existence of a $M=2.0~\msun$ neutron star.}
{We consider a  non-linear relativistic mean field (RMF) model involving baryon
octet coupled to meson fields. An effective lagrangian includes quartic terms in
the meson fields. The values of the parameters of the model are obtained by fitting
semi-empirical parameters of nuclear matter at the saturation point, as well as
potential wells  for hyperons in nuclear matter and the strength of the
$\Lambda-\Lambda$ attraction in double-$\Lambda$ hypernuclei.}
{We propose a non-linear  RMF model which is consistent with up-to-date semiempirical
nuclear and hypernuclear data and allows for neutron stars with hyperon cores and
$M>2~\msun$.  The model involves hidden-strangenes scalar and vector mesons, coupled to hyperons only,
and quartic terms involving  vector meson fields. }
{Our  EOS involving hyperons is
stiffer than the corresponding nucleonic
EOS (with hyperons artificially suppressed) above
five  nuclear densities. Required stiffening is generated by the quartic
terms involving hidden-strangeness vector meson. }


\titlerunning{$2\;{\rm M}_\odot$ pulsar and hyperon cores}
\authorrunning{}
\maketitle

\section{Introduction}
\label{sect:introduction}

Recent measurement of the mass of PSR J1614-2230, $1.97\pm 0.04~{\rm M}_\odot$
(Demorest et al. 2010), puts stringent constraint on the equation of state (EOS)
 of dense matter in neutron star cores. In the light of this measurement,
 EOSs of dense matter, based on the  modern many-body theories and realistic
 strong-interaction model, lead to a puzzle. On the one hand, interactions
 consistent with experimental data on hypernuclei, lead to the presence
  of hyperons at the densities exceeding $2-3\rho_0$, where
 $\rho_0=2.7\times 10^{14}~\mdens$  (corresponding to the baryon number
density $n_0=0.16~\bdens$) is the normal
 nuclear density. On the other hand, inevitable softening of the
EOS, due to the hyperonization, implies
 the maximum allowable mass $M_{\rm max}\lesssim 1.5~\msun$ (see, e.g.
\citealt{Burgio2011,Vidana2011}, and references therein). Such a low $M_{\rm max}$
is only marginally consistent with $1.44~\msun$ of the Hulse-Taylor pulsar, but was
contradicted already  by  $1.67\pm 0.04~\msun$ of PSR J1903-0327 (Champion et al.
2007;  more precise value has been recently obtained by Freire et al. 2010). Let us
mention, that this problem  cannot be solved by adding an ad hoc extremely stiff
repulsive three-body contribution to the EOS \citep{Vidana2011}.

We consider neutron star cores composed of baryons, electrons, and muons. Baryons
and leptons are in weak-interaction equilibrium. Hyperons appear at density
$\rho_1$ (baryon density $n_1$). For $\rho<\rho_1$ only nucleons are present. For
$\rho>\rho_1$  matter contains a mixture of nucleons and hyperons. This state
(phase) will be denoted as NH. We will also consider a dense matter model with
hyperons artificially suppressed. Such a purely nucleon state will be denoted as N.
The corresponding EOS will be denoted as EOS.NH and EOS.N. These EOS  coincide for
$\rho<\rho_1$.

A too low $M^{\rm(NH)}_{\rm max}$ is not an inevitable feature
of neutron stars with hyperon cores. In a very recent paper \cite{Bonanno2011}
get $M^{\rm(NH)}_{\rm max}>2~\msun$, starting from an extremely stiff
relativistic mean field model NL3 EOS.N, which yields
$M^{\rm(N)}_{\rm max}=2.8~\msun$ (close to the absolute upper
bound on $M_{\rm max}$  stemming from causality,
e.g., \citealt{NSbook2007}).   \cite{Bonanno2011} extend the NL3 model to
the hyperon sector, and get $M^{\rm(NH)}_{\rm max}=2.03~\msun$. Massive
stars with hyperon core exist there
because of the extreme stiffness of EOS.N, obtained at the
cost of having  a parameter of nuclear-matter  parameter (the
slope of symmetry energy vs. density)  $L=118~$MeV  significantly higher than
its semi-empirical estimates.

In our approach we keep $L$ (and the stiffness of the EOS.N near $\rho_0$)
 within the semi-empirical (i.e., obtained within a model of atomic nuclei,
 and therefore model-dependent) estimates. Our EOS.N at high density is stiff,
but  not extremely stiff: $M^{\rm(N)}_{\rm max}=2.1~\msun$. Generally,
$M^{\rm(NH)}_{\rm max}$  is essentially determined by the $\rho\ga 5\rho_0$ segment
of the EOS.NH. So if the hyperon softening occurs at $2-3\rho_0$, then, to get a
sufficiently large $M_{\rm max}$, the softening  must be followed by a sufficiently
strong {\it stiffening} of EOS.NH  for $\rho\ga 5\rho_0$. One has therefore to find
a mechanism of stiffening of EOS.NH at these densities. For our model of EOS.N, in
order to yield $M^{\rm (NH)}_{\rm max}>2~\msun$,
 EOS.NH for $\rho\ga 5\rho_0$ should  be actually {\it stiffer} than
the EOS.N one. Simultaneously, the NH phase has to be stable (thermodynamically
preferred over the N one). We derive a constraint on the EOS.NH resulting
from the conditions mentioned above, and discuss consequences of violation of this
constraint.

In our discussion we restrict to hadronic matter, and do not consider a possibility
of quark deconfinement. Because of the surface effects and electrical screening in
a quark plasma, transition to a quark matter would  occur  at nearly constant
pressure and with an only  slightly smoothed density jump (see, e.g.,
\citealt{Endo2006}, and references therein). A reasonable scenario is: softening of
the N phase by hyperonization at $2-3\rho_0$, then softening of the NH phase by
quark deconfinement at a significantly higher density. \cite{Bonanno2011} show that
assuming NL3 EOS.NH, a transition to quark matter occurring after hyperonization
could be consistent with $2~\msun$ pulsar provided the vector repulsion in quark
matter is sufficiently strong and quark deconfinement takes place near the maximum
NS mass. For our EOS.NH getting $2~\msun$ with a quark core would require a very
fine tuning, and we do not consider such an unlikely possibility.

The problem of an interplay of attraction (softening) and
repulsion (stiffening) in dense hadronic matter can be formulated in simple
terms using a modern  effective field theory,
involving baryon and meson fields. In the case of nucleon
matter,  such a theory can be put on the firm theoretical basis, starting from the
 QCD (\citealt{Walecka2004}, and references therein).
 Such an effective theory can
be solvable  within the mean field approximation and can give a satisfactory
description of a wealth of nuclear physics data if the coupling of nucleons to the
three meson fields: scalar $\sigma$, vector $\omega_\mu$ and $\rho_\mu^i$ , is
considered. Here, $\mu$ and $i$ denote the space-time and isospin-space components
of the field. Effective lagrangian contains  quadratic and quartic terms in vector
fields, and quadratic, cubic, and quartic terms in scalar fields. While $\sigma$
yields attraction to bind nuclei, vector meson fields generate repulsion to
saturate nuclear matter at $\rho_0$. Numerical coefficients in the effective
lagrangian are fixed by fitting a wealth of nuclear data
(\citealt{SugaharaToki1994}). The effective model is then extended to include the
hyperon sector. Two meson fields with "hidden strangeness" ($\overline{s}s$) are
added: scalar $\sigma^*$ (quadratic terms) and vector-isovector $\phi_\mu^i$
(quadratic and quartic terms). These fields couple to hyperons only
(\citealt{SchaffnerDover1994}). An important constraint on the hyperon sector of
lagrangian results from the existing evaluations of the depth of the potential well
acting on a single zero-momentum hyperon in nuclear matter, $U^{(N)}_\Lambda$,
$U^{(N)}_\Sigma$,and $U^{(N)}_\Xi$ (binding energy of a hyperon in nuclear
matter is $B_H=-U^{(N)}_H$). An effective theory of hadronic matter described in general
terms above and solved in the mean field approximation will be referred to as
non-linear relativistic mean field model (non-linear RMF, \citealt{BednarekM2009}).

There exist a few older and simpler models of NH matter, yielding $M^{\rm(NH)}_{\rm
max}>2~\msun$, but they reach this aim by  pushing the parameters of the
interaction  to the extreme and likely unrealistic  limits. These models are based
on the relativistic mean field lagrangian involving octet of baryons coupled to
$\sigma$, $\omega_\mu$ and $\rho_\mu^i$  meson fields (quadratic terms in the
lagrangian), with additional cubic and quartic $\sigma$ self-interaction terms (for
a review, see \citealt{GlendBook}). The mean field solutions of the field equations
are called relativistic mean field (RMF) model.
 \cite{GlendMoszk1991} can exceed $2~\msun$
assuming unrealistically high  nuclear matter incompressibility, $K=300~$MeV, and
an unrealistically strong $\Lambda-\sigma$ attraction, balanced by a $\Lambda-\omega$
repulsion to get experimental $U^{(N)}_\Lambda$. Relations between coupling constants
$g_{H\omega}$ and $g_{N\omega}$, resulting from the SU(6) symmetry and making them
significantly weaker than $g_{N\omega}$, have to be strongly violated. Finally,
experimental constraints on  $U_\Sigma$ and $U_\Xi$ in nuclear matter are not
applied.  Instead, it is assumed that all hyperons in the baryon octet have the
same coupling as $\Lambda$. Similar model was used to obtain $M^{\rm (NH)}_{\rm
max}>2~\msun$ by \cite{Bombaci2008}.  Very recently,  the RMF model was applied to
calculate $M^{\rm (NH)}_{\rm max}$ under the conditions $U^{(N)}_\Lambda=-30$~MeV,
$U^{(N)}_\Sigma=+30~$MeV, $U^{(N)}_\Xi=-18~$MeV and  varying $g_{H\omega}$
(\citealt{Debarati2011}). For the SU(6)-symmetric set of $g_{B\omega}$,
the maximum mass is about $1.7~\msun$.  $M^{\rm (NH)}_{\rm max}>2~\msun$
can be obtained only for sufficiently strong
 $g_{H\omega}\simeq g_{N\omega}$, i.e. by breaking the SU(6) symmetry.

In the present paper we propose a solution of the puzzle "hyperonization $-$
$M_{\rm max}>2~\msun$" using a specific realization of the non-linear RMF model of
hadronic matter (\citealt{BednarekM2009}). The  non-linear RMF
model of NH matter is presented in Sect.\ref{sect:BM}.  Experimental
constraints  from nuclear and hypernuclear physics are described in
Sect.\ref{sect:BM-parameters}. A particular
EOS.NH is described in Sect.\ref{sect:EOS}. A model of PSR J1614-2230 with
a hyperon core is presented in Sect.\ref{sect:1.97}. The problem of high-density
instability of the NH phase and the $M-R$ relation for neutron stars
models are  discussed in Sect.\ref{sect:NH.H}.  Finally, in
Sect.\ref{sect:conclusions} we summarize results of the paper,
compare them with results obtained by other authors,
 and present our conclusions.

Preliminary results of our work were presented at  the  MODE-SNR-PWN Workshop in
Bordeaux, France, November 15-17, 2010, and  in a poster at
the CompStar 2011 Workshop in Catania, Italy, May 9-12, 2011.

\section{Non-linear RMF model of hyperon cores}
\label{sect:BM}
The model has been formulated by \cite{BednarekM2009}. The octet of
baryons includes nucleon doublet $N$ and six lowest-mass hyperons $H$: $\Lambda$
singlet, $\Sigma$ triplet and $\Xi$ doublet. Uniform number density of each baryon
species $B$ is denoted $n_B$ ($B=n,p,\Lambda,\ldots$).

In the nucleon sector the meson fields are: scalar $\sigma$, vector $\omega_\mu$, and
vector-isovector  $\rho^i_\mu$. Generalization of the non-linear RMF model to the
baryon octet is done in the following way. Additional "hidden-strangeness"
 mesons: scalar $\sigma^{_\star}$ and  vector $\phi_\mu$ are introduced.
They  couple to hyperons only, $g_{N\sigma^{_{\star}}}=g_{N\phi}=0$. The
vector-meson coupling constants to hyperons are assumed to fulfill relations
stemming from the SU(6) symmetry (additive quark model):

\begin{eqnarray}
g_{\Lambda\omega}=g_{\Sigma\omega}=2g_{\Xi\omega}
=\frac{2}{3}g_{N\omega}~,~~g_{\Sigma\rho}=2g_{\Xi\rho}=2g_{N\rho}~, \\
\nonumber
g_{\Lambda\phi}
=g_{\Sigma\phi}=\frac{g_{\Xi\phi}}{2}
=\frac{\sqrt{2}}{3}g_{N\omega}.
\label{vec-coup}
\end{eqnarray}
Similar symmetry relations can be obtained for the coupling constants
of the scalar mesons,  but they are not used in the present
model. Instead, we are adjusting them to fit experimental estimates
of $U^{(N)}_B$.

At fixed $\lbrace n_B\rbrace$ $(B=n,p,\Lambda,\dots)$, and assuming vanishing
baryon currents, the hadronic lagrangian density ${\cal L}_{\rm had}$ is used to
derive equations of motion for the meson fields. Static solutions are found
assuming that baryonic matter is isotropic and uniform. Mean-field approximation,
neglecting quantum corrections, is used: $\sigma\longrightarrow \langle
\sigma\rangle=s_0$, $\omega_\mu\longrightarrow
\langle\omega_\mu\rangle=w_0\delta_{\mu0}$, $\rho^i_\mu\longrightarrow
\langle\rho^i_\mu\rangle=r_0\delta_{\mu0}\delta_{i3}$,
$\sigma^{_\star}\longrightarrow \langle\sigma^{_\star}\rangle=s^{_\star}_0$,
$\phi_\mu\longrightarrow \langle\phi_\mu\rangle=f_0\delta_{\mu0}$. The resulting
lagrangian density function ${\cal L}_{\rm had}$ consists of  three components,
${\cal L}_{\rm had}={\cal L}_{\rm B}+{\cal L}_{\rm M}^{(2)}+ {\cal L}_{\rm
M}^{(3,4)}$.  ${\cal L}_{\rm B}$ is obtained from the free-baryon lagrangian by
replacing bare baryon masses $m_B$ ($B=n,p,\Lambda~,\ldots$) by the effective ones,
$m^\star_B=m_B-g_{B\sigma}s_0-g_{B\sigma^{_\star}}s^\star_0$. The quadratic
(interaction) component ${\cal L}_{\rm M}^{(2)}$ contains terms proportional to
$s^2_0,~w^2_0,r^2_0, {s^{\star}_0}^ 2,f^2_0$. The interaction component ${\cal
L}_{\rm M}^{(3,4)}$ contains terms cubic and quartic in $s_0$, and quartic
vector-meson terms proportional to $w_0^4, r_0^4,f_0^4$ and the cross terms
proportional to $f_0^2 w_0^2$, $f_0^2r_0^2$ and $w_0^2r_0^2$.

The hadronic lagrangian density function
${\cal L}_{\rm had}$ is then used to calculate the
hadron energy-density as a function of partial baryon densities $\lbrace
n_B\rbrace$, ${\cal E}_{\rm had}(\lbrace n_B\rbrace)$.
Calculations done for the considered model
(\citealt{BednarekM2009}) lead to the following
explicit formulae for the hadron contribution to the
energy density ${\cal E}$ and pressure $P$ (notice that original equations
in \cite{BednarekM2009}
 contain several misprints which are corrected below;
 for the sake of simplicity, we use
 a shorthand notation $g_{N\sigma}\equiv g_{\sigma}~,
g_{N\omega}\equiv g_{\omega}~,\dots$)
\begin{eqnarray}
\nonumber
&{\cal E}_{\rm had}&=
+\frac{1}{2}m_{\sigma}^{2}s_{0}^{2}
+\frac{1}{2}m_{\omega}^{2}w_{0}^{2}
+\frac{1}{2}m_{\rho}^{2}r_{0}^{2}+ U(s_{0})
\\
\nonumber
&~&+\sum_{B}\frac{2}{\pi^{2}}\int_{0}^{k_{F,B}}k^{2}dk\sqrt{k^{2}
+(m_{B}-g_{\sigma B}s_{0}\boxed{-g_{\sigma^{\ast}}s_{0}^{\ast}})^{2}}
\\
\nonumber
& &+3\Lambda_{V}(g_{\rho}g_{\omega})^{2}w_{0}^{2}r_{0}^{2}
+\frac{3}{4}c_{3}\left(w_{0}^{4}+r_{0}^{4}\right)
\\
\nonumber
&~& \boxed{+\frac{1}{2}m_{\phi}^{2}f_{0}^{2}
+\frac{1}{2}m_{\sigma^{\ast}}^{2}s_{0}^{\ast 2}}
+\boxed{3\left(\frac{1}{8}c_{3}+\frac{1}{4}\Lambda_{V}(g_{\rho}g_{\omega})^{2}
\right)f_{0}^{4}}
\\
&~&\boxed{+3\left(
\frac{3}{4}c_{3}-\frac{1}{2}\Lambda_{V}(g_{\rho}g_{\omega})^{2}\right)
f_{0}^{2}(w_{0}^{2}+r_{0}^{2})}~,
\label{eq:E_had}
\end{eqnarray}
\begin{eqnarray}
\nonumber
&P_{\rm had}&=-\frac{1}{2}m_{\sigma}^{2}s_{0}^{2}+
\frac{1}{2}m_{\rho}r_{0}^{2}+\frac{1}{2}m_{\omega}w_{0}^{2}
-U(s_{0})
\\
\nonumber
&+&\sum_{B}\frac{1}{3\pi^{2}}\int_{0}^{k_{F,B}}
\frac{k^{4}{\rm d}k}{\sqrt{(k^{2}+m_{B}-g_{\sigma B}s_{0}
\boxed{-g_{\sigma^{\ast}}s_{0}^{\ast}})^{2}}}
\\
\nonumber
&~&+
\frac{1}{4}c_{3}(w_{0}^{4}+r_{0}^{4})
+\Lambda_{V}(g_{\rho}g_{\omega})^{2}w_{0}^{2}r_{0}^{2}
\\
\nonumber
&~&\boxed{+\frac{1}{2}m_{\phi}^{2}f_{0}^{2}
-\frac{1}{2}m_{\sigma^{\ast}}^{2}s_{0}^{\ast 2}}
+\boxed{\left(\frac{1}{8}c_{3}+\frac{1}{4}\Lambda_{V}(g_{\rho}g_{\omega})^{2}
\right)f_{0}^{4}}\\
&~&\boxed{+\left(\frac{3}{4}c_{3}-\frac{1}{2}\Lambda_{V}(g_{\rho}g_{\omega})^{2}\right)
f_{0}^{2}(w_{0}^{2}+r_{0}^{2})}~,
\label{eq:P_had}
\end{eqnarray}
\vskip 3mm
where the non-linear $\sigma$-self-interaction potential is
\begin{equation}
U(\sigma)=-\frac{\kappa}{3!}\sigma^3-\frac{\lambda}{4!}\sigma^4~.
\label{eq:U.sigma}
\end{equation}
The terms vanishing in purely nucleon (zero strangeness) matter are put into the
rectangles.

The   quartic terms in ${\cal E}_{\rm had}$ and $P_{\rm had}$ deserve
additional explanations. Their form stems from the chiral SU(3)
symmetry of the baryon-meson and meson-meson interactions.
The  coefficients of the quartic terms involve two phenomenological parameters,
$c_3$ and $\Lambda_V$.

Let us consider first the quartic terms in the nucleon sector. The
vector-isoscalar quartic term ($w_0^4$) was already present in the TM1
model of \cite{SugaharaToki1994}. However, TM1 was constructed to describe
atomic nuclei, and therefore it  is valid for nuclear matter near saturation
density and for small neutron excess. It is to be expected that
extrapolation to supranuclear density and large neutron excess
necessitates a richer  isospin and
density dependence of the model lagrangian than that assumed in the
TM1 model. \cite{BednarekM2009} proposed to do this by enlarging the
quartic terms via adding a vector-isovector one ($r_0^4$) and a
cross-term ($w_0^2r_0^2$). The strengths of the quartic terms are
determined by two parameters, $\Lambda_V$ and $c_3$, instead of only
one in the TM1 model of \cite{SugaharaToki1994}. This allows for
a good fitting not only (semi-empirical estimates of) nuclear symmetry
energy and incompressibility, but also the slope parameter $L$, and
simultaneously yields $M_{\rm max}>1.97\;\msun$.

As shown in (\citealt{BednarekM2009}), chiral SU(3) symmetry
yields a suitable extension of the quartic terms to the hyperon sector,
the same $c_3$ and $\Lambda_V$ entering the quartic-terms coefficients.
Additional quartic terms in the hyperon sector are generated by the
 hidden-strangeness vector-isoscalar  field $f_0$.
\begin{table}
\begin{center}
\caption{BM165 model of hadronic matter.
Calculated  nuclear matter parameters at the
saturation point}
\begin{tabular}[t]{ccccc}
\hline\hline
&&&&\\
 $n_{\rm s}$  & $E_{\rm s}$   & $K_{\rm s}$ & $S_{\rm s}$ &   $L$ \\
 $({\rm fm^{-3}})$  & (MeV)   & (MeV)& (MeV) & (MeV)\\
 \hline
 &&&&\\
  0.145   & -16.3  & $279$ &    $33$
  &  74 \\
    &&&&\\
 \hline\hline
\end{tabular}
\end{center}
\end{table}

\begin{table}
\begin{center}
\caption{BM165 model of hadronic matter. Calculated  zero-momentum  single-baryon
potentials in symmetric nuclear matter, $U^{{\rm (N)}}_B$ and of  $\Lambda$ in
$\Lambda$ matter, $U^{(\Lambda)}_\Lambda$. All results are obtained at $n_{\rm
b}=n_{\rm s}$}
\begin{tabular}[t]{ccccc}
\hline\hline
&&&&\\
  $U^{{(N)}}_N$  &  $U^{({N})}_\Lambda$ &
   $U^{({N})}_\Sigma$& $U^{({N})}_\Xi$ &
   $U^{(\Lambda)}_{\Lambda}$\\
 (MeV)  & (MeV)   &  (MeV)& (MeV) & (MeV) \\
  & &&&\\
  \hline
  &    &   & &\\
      -69 &  -28    &  +30 &  -18  &  -5\\
&&&&\\
\hline\hline
\end{tabular}
\end{center}
\end{table}


\section{Determination of parameters of non-linear RMF  model}
\label{sect:BM-parameters}
Let us denote the neutron excess in nuclear matter by $\delta =(n_n-n_p)/n_{\rm
b}$. The energy per nucleon (excluding nucleon rest energy) is $E(n_{\rm
b},\delta)$.  Analysis of a wealth of data on heavy atomic nuclei can yield the
parameters of nuclear matter near the saturation point, corresponding to the
minimum of energy per nucleon, $E_{\rm s}$,  reached at $n_{\rm b}=n_{\rm s}$ and
$\delta=0$. Results are model-dependent and therefore they are called
semi-empirical. Other semi-empirical parameters are: symmetry energy $S_{\rm s}$,
incompressibility $K_{\rm s}$,  and the symmetry-energy slope
parameter $L$,

\begin{eqnarray}
\nonumber
S_{\rm s}&=&\left(\frac{\partial^2 E}{\partial \delta^2}
\right)_{n_{\rm s},\delta=0}~,~
K_{\rm s}=
9n_{\rm s}^2\left(\frac{\partial^2 E}{\partial n_{\rm b}^2}
\right)_{n_{\rm s},\delta=0}~,\\
\nonumber
\\
L&=& 3 n_{\rm s}\left(\frac{\partial^3 E}
{\partial n_{\rm b}\partial \delta^2}\right)_{n_{\rm s},\delta=0}~.
\label{eq:NMatt.param}
\end{eqnarray}

Studies of hypernuclei and of $\Sigma^-$ atoms allow for evaluation of the
potential energy of a {\it single}  zero momentum hyperon in symmetric nuclear matter,
$U^{(N)}_H$. The non-linear RMF yields following expression for this
quantity:
\begin{equation}
U^{(N)}_H=g_{H\sigma}s_0-g_{H\omega}w_0~,
\label{eq:U.H}
\end{equation}
which  should be calculated at $n_{\rm s}$ and $\delta=0$.
The semi-empirical estimates are $U^{(N)}_\Lambda=-28~$MeV,
$U^{(N)}_\Sigma=+30~$MeV, $U^{(N)}_\Xi=-18~$MeV
(\citealt{SchaffnerGal2000}).
Eq.\;(\ref{eq:U.H}) is then used to determine $g_{\Lambda\sigma}$,
$g_{\Sigma\sigma}$ and $g_{\Xi\sigma}$.

As we are dealing with NH phase, containing
 {\it finite} fractions of hyperons,
we need information on the hyperon-hyperon interaction. Studies of
double-$\Lambda$ hypernuclei suggests that $\Lambda-\Lambda$ interaction
is attractive. In the mean-field approximation it can be
characterized by the potential well of a zero momentum
$\Lambda$ in the $\Lambda$-matter. In our model we get a general
expression for a potential energy of a zero-momentum hyperon $H^\prime$
in the $H$-matter,

\begin{equation}
U^{(H^\prime)}_H=g_{H\sigma}s_0-g_{H\omega}w_0+
g_{H\sigma^{_\star}}s_0^{_\star}-g_{H\phi}f_0~.
\label{eq:U.HH}
\end{equation}

The latest (very uncertain) semi-empirical estimate coming from double-$\Lambda$
hypernuclei  is $U^{(\Lambda)}_\Lambda=-5~$MeV
(\citealt{Takahashi2001,SongSu2003}).
Eq.\;(\ref{eq:U.HH}) is then
used to determine $g_{\Lambda{\sigma^{_\star}}}$. For $U^{(\Sigma)}_\Sigma$
and  $U^{(\Xi)}_\Xi$ no data  exist. Therefore, we
estimate them using  the relations
\begin{equation}
U^{(\Xi)}_\Xi\simeq U^{(\Xi)}_\Lambda \simeq 2U^{(\Lambda)}_\Xi\simeq
2U^{(\Lambda)}_\Lambda~.
\label{eq:relations.U.HH1}
\end{equation}
These relations have been established on the basis of one-boson exchange models
and semi-empirical evaluation of the
 strength of the $\Lambda-\Lambda$ attraction (\citealt{SchaffnerDover1994}).

We adjusted parameters of our lagrangian to reproduce,  within a few percent,
ten semi-empirical nuclear and hyper-nuclear data. Adjusting
 the isovector parameters in the lagrangian density, $g_\rho$ and $\Lambda_V$,
deserves an additional explanation. This was done by employing existing
information on the density dependence of the
symmetry energy. Namely, we used not only the value of
the symmetry energy at saturation, $S_{\rm s}$, but also a semi-empirical estimate of
symmetry energy at $n_{\rm b}\approx 0.1~{\rm fm}^{-3}$, $26.67~$MeV, which
determined the value of $L$ (\citealt{HorowitzPiekarewicz2001}). This influenced
the EOS of neutron matter,
because $E(n_{\rm b},1)\approx E(n_{\rm b},0.5)+S(n_{\rm b})$ (see, e.g.,
\citealt{NSbook2007})
\begin{figure}[h]
\resizebox{\columnwidth}{!}
{\includegraphics[angle=-90,clip]{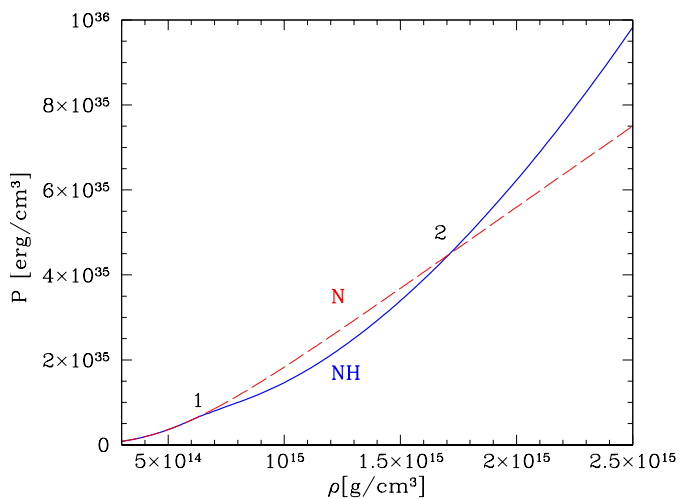}}
\caption{Equations of state EOS.N and EOS.NH calculated
using the BM165 model. Hyperons appear at point 1 and EOS.NH crosses the EOS.N
one at point 2.
}
\label{fig:Prho}
\end{figure}

\begin{figure}[h]
\resizebox{\columnwidth}{!} {\includegraphics[angle=-90,clip]{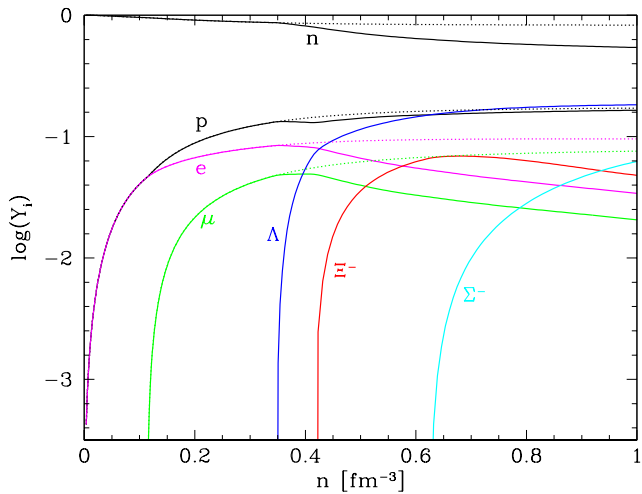}}
\caption{Number fractions of the constituents of dense matter in beta equilibrium,
$Y_{\rm i}=n_{\rm i}/n_{\rm b}$, versus baryon density, $n_{\rm b}$. Dotted lines:
EOS.N. Solid lines: EOS.NH. } \label{fig:yin_165}
\end{figure}

\section{EOS of neutron-star matter}
\label{sect:EOS}
Total energy density  and total pressure are  sums of contributions of hadrons and
leptons, because contributions of electromagnetic interaction to these quantities
are negligibly small, and leptons (electrons and muons)
 can be treated as ideal Fermi gases,
\begin{equation}
{\cal E}={\cal E}_{\rm had}+{\cal E}_{\rm lep}~,~~
P=P_{\rm had}+P_{\rm lep}~.
 \label{eq:E.components}
\end{equation}
Let us fix the (total) baryon number density,
\begin{equation}
n_{\rm b}=\sum_B n_B~.
\label{eq:n.b}
\end{equation}
Imposing electrical charge neutrality and beta
equilibrium, one gets a system of non-linear equations for the
particle species fractions $Y_i=n_i/n_{\rm b}$. Let at a given $n_{\rm b}$
the equilibrium fractions be $\lbrace Y^0_i(n_{\rm b})\rbrace$. Under
imposed conditions, energy density and pressure become
 functions of $n_{\rm b}$ only.
For the sake of comparison, we will consider not only a general EOS involving
nucleons and hyperons, EOS.NH, but also an EOS of nucleon matter, EOS.N, with
artificially suppressed hyperons (this was done by doubling the
actual hyperon rest masses).

Expression for $P_{\rm had}$ (Eq.\ref{eq:P_had}) shows that hyperons produce
new repulsive quartic terms involving $f_0$, $w_0$,  and $r_0$. The dependence of
the EOS on $\Lambda_V$ turns out to be  quite strong. All other parameters being determined by the conditions of
reproducing nuclear and hypernuclear data (i.e., near the saturation point of nuclear
matter), the high-density stiffness of EOS.NH increases monotonically
with increasing $\Lambda_V$. In what follows we will use
  $\Lambda_V=0.0165$, which as we will see yields  a high-density stiffness
  of the EOS.NH consistent   with $M_{\rm max}^{\rm (NH)}>2~\msun$, while
  keeping good agreement with semi-empirical nuclear matter parameters.
  This EOS.NH will be hereafter   referred to as BM165.NH or BM165.
Thermodynamical equilibrium in the NH matter imposes
relations between  chemical potentials of hyperons {\it present}
in dense matter, nucleons, and leptons. These relations can be
expressed in the general form (see, e.g., \citealt{NSbook2007}):
\begin{equation}
\mu_{_H}=\mu_n-q_{_H}\mu_e~,~~~~~\mu_e=\mu_\mu~,
\label{eq:NH.eq}
\end{equation}
where $q_{_H}$ is the charge of  hyperon  $H$ in units
of the proton charge. The threshold density  for appearance
of hyperons $H$, $n_{\rm b}^{H}$, is determined by the enthalpy of a single
hyperon $H$ in a beta-equilibrated dense matter,
\begin{equation}
\mu_H^{(0)}=m_{_H}c^2+U_H+P/n_{\rm b}~,
\label{eq:mu.thr}
\end{equation}
where $U_H$ is the potential energy of this hyperon. For
$n_{\rm b}<n^{(H)}_{\rm b}$, $\mu^{(0)}_H>\mu_n-q_{_H}\mu_e$,
and $H$ decays in a weak interaction process. At the threshold density
a single $H$ in  dense matter is stable,
\begin{equation}
\mu_H^{(0)}=\mu_n-q_{_{H}}\mu_e~,
\label{eq:NH.thr}
\end{equation}
and for  $n_{\rm b}>n^{H}_{\rm b}$ the density of stable $H$ grows
with increasing $n_{\rm b}$.

 For the BM165 model first hyperon
to appear is $\Lambda$ at $6.3\times
10^{14}~\mdens$ ($0.35~\bdens$), where
$\mu_H^{(0)}=\mu_n$. Next hyperon is $\Xi^-$,
it appears at $7.7\times 10^{14}~\mdens$
($0.42~\bdens$ ), where  $\mu_{\Xi^-}^{(0)}=
\mu_n+\mu_e$. Repulsive $U^{(N)}_{\Sigma^-}$
implies that $\Sigma^-$ appears at significantly
higher density than $\Xi^-$, namely $1.23\times 10^{15}~\mdens$
($0.63~\bdens$), in spite of $m_{\Sigma^-}<m_{\Xi^-}$.

The order of appearance of hyperons in dense matter deserves
a comment. For a long time, in view of  lack of an experimental
information on $U^{(N)}_\Sigma$ and $U^{(N)}_\Xi$, they were
assumed to be similar to $U^{(N)}_\Lambda$. Consequently, $\Sigma^-$
was found to be the first hyperon to appear, not the lightest hyperon
$\Lambda$, because the (unfavorable) effect of $m_{\Sigma^-}>m_\Lambda$
was weaker than the (favorable) effect of the presence of $\mu_e$ in the
threshold condition $\mu_{\Sigma^-}^{(0)}=\mu_n+\mu_e$ (see, e.g.,
\citealt{NSbook2007}). However, large positive (repulsive) $U^{(N)}_\Sigma$
resulting from analyses of the $\Sigma^-$-atoms pushes
$n^{\Sigma^-}_{\rm b}$ well above $n^{\Xi^-}_{\rm b}$.
Consequently, $\Sigma^-$ is the last,  instead of being the first,
 to appear in neutron star core.

Appearance of hyperons leads to significant softening
of the EOS.NH compared to EOS.N
(Fig. 2). In order to support neutron stars
with $M>2\;\msun$, the EOS.NH  has {\it necessarily} to
significantly {\it stiffen}
at higher densities. The curve $P^{\rm NH}(\rho)$ crosses
the $P^{\rm N}(\rho)$ one
at $\rho_2=1.76\times 10^{14}~\mdens$  ($n_2=0.85~\bdens$),
and at higher density EOS.NH is stiffer than EOS.N.

Actually,  the difference  $P^{\rm NH}-P^{\rm N}$ is limited by the
condition  of stability of the NH phase against the re-conversion into the N
phase.  Assume that matter is in beta
equilibrium. At $T=0$ the small change
of energy per baryon  ${\rm d}E$ is related to the small change
of baryon density ${\rm d}n_{\rm b}$ by
\begin{equation}
{\rm d}E=P{{\rm d}n_{\rm b}\over n_{\rm b}^2}~.
\label{eq:dE.dn_b}
\end{equation}
Therefore, condition of stability of the NH phase
(against the conversion into the N one)
\begin{equation}
E^{\rm NH}(n_{\rm
b})<E^{\rm N}(n_{\rm b})~,
\label{eq:stab.NH.N}
\end{equation}
implies
\begin{equation}
\int_{n_1}^{n_{\rm b}}{{P^{\rm NH}(n^\prime_{\rm b})-
P^{\rm N}(n^\prime_{\rm b})}\over {n^\prime_{\rm b}}^2}
{\rm d}n^\prime_{\rm b}<0~,
\label{eq:N.NH.ineq}
\end{equation}
where $n_1$ is the density at the  first hyperon threshold. For the BM165 model
$n_1=0.35~\bdens$.

\begin{figure}[h]
\resizebox{\columnwidth}{!} {\includegraphics[angle=-90,clip]{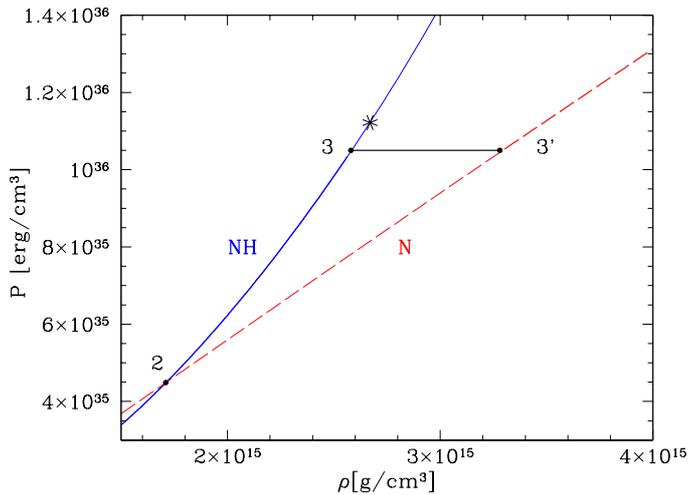}}
\caption{EOS BM165 in the
vicinity of the high-density NH-N phase transition. 2: crossing of the
the N and NH pressures. 3: density and pressure of the NH phase at the
phase coexistence. $3^\prime$: density and pressure of the N phase at the
phase coexistence. Horizontal segment $33^\prime$ - pressure at the 1-st
order phase transition. Were the NH phase be (sufficiently) stable
till the maximum mass, maximum (central) density in stable stars would correspond
to the asterisk sign, $\rho_{\rm c,max}=\rho_\ast$. For  the fully equilibrated
hadronic matter $\rho_{\rm c,max}$ is slightly higher than $\rho_3$.
For a more detailed  discussion of this point see the text.
 }
\label{fig:Prho.NH.N}
\end{figure}
\begin{figure}[h]
\resizebox{\columnwidth}{!} {\includegraphics[angle=-90,clip]{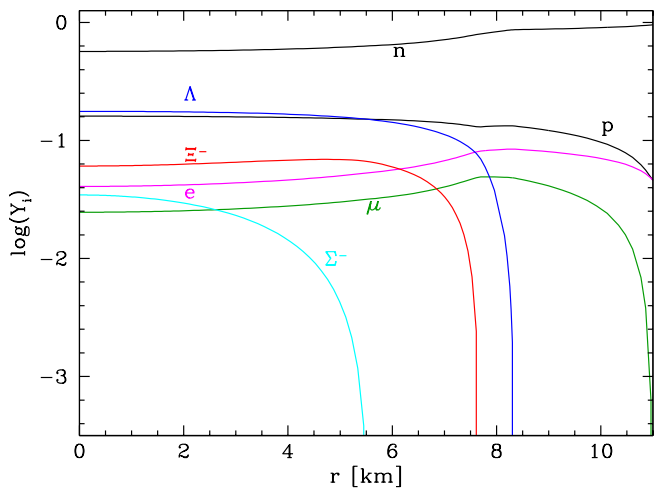}}
\caption{The logarithm of the number fractions of the constituents of dense matter,
${\rm log}_{10}(Y_{\rm i})$, versus  circumferential radius, in
the liquid core of a $1.97~\msun$ star model based on the BM165 EOS. }
\label{fig:Yi_1.97}
\end{figure}

\section{A model of PSR J1614-2230}
\label{sect:1.97}
To get a complete EOS of the neutron-star interior, the
 BM165 EOS of the liquid core was supplemented with an
 EOS of the crust. We used the EOS of the inner crust
  of \cite{DouchinH2001}, the model of \cite{HaenselP1994} for the
outer crust down to $10^8~\mdens$, and the classical model of \cite{BPS1971} for
the outer layer with $\rho<10^{8}~\mdens$. A model of neutron star of gravitational
mass $1.97~\msun$, rotating rigidly at $317~$Hz, was calculated using the 2-D {\tt
rotstar} code from the LORENE library ({\tt http://www.lorene.obspm.fr}) implementing
the formulation of (\citealt{BGSM1993}). The
circumferential equatorial radius of neutron star
is $R_{\rm eq}=11.83~$km, central density
$\rho_{\rm c}=1.73\times 10^{14}~ \mdens$ and central baryon density $n_{\rm
c}=0.834~\bdens$. At 317 Hz, polar flattening  is rather small: the  radial
coordinate at the equator is only $200~$m larger the that at the pole. The number
fractions of the particle species $Y_i=n_i/n_{\rm b}$, versus radial coordinate
$r$, in the liquid core of neutron star, are plotted in Fig.\;\ref{fig:Yi_1.97}.
The radius of the hyperon core is $8.36~$km. The strangeness per baryon at the
star's center
 $(S/N_{\rm b})=-0.35$.
%
\section{High-density instability of the NH phase and neutron star models}
\label{sect:NH.H}
Violation of inequality (\ref{eq:N.NH.ineq}) indicates instability of the NH phase
against it conversion into the purely nucleon (N) one. Thermodynamic equilibrium of
dense matter at pressure $P$ corresponds to the minimum  of baryon chemical
potential $\mu_{\rm b}=({\cal E}+P)/n_{\rm b}$. An equilibrium phase-transition
NH$\longrightarrow$N occurs at $P_3$ such that
\begin{equation}
\mu_{\rm b}^{\rm (NH)}(P_3)=\mu_{\rm b}^{\rm (N)}(P_3)~,
\label{eq:NH.H}
\end{equation}
and is accompanied by a density jump from $\rho_3=\rho^{\rm (NH)}(P_3)
=2.58\times 10^{15}~\mdens$ ($n_3=1.105~\bdens$) on the NH
side to $\rho^\prime_3=\rho^{\rm (N)}(P_3)=3.25\times 10^{15}~\mdens$
($n^\prime_3=1.31~\bdens$) on the N side. The BM165 EOS in the
vicinity of  $P_3=1.05\times 10^{36}~{\rm erg~cm^{-3}}$ is shown in
Fig.\;\ref{fig:Prho.NH.N}. The softening of the EOS for $P>P_3$ is twofold. First,
there is a constant pressure sector of the EOS (vanishing compression modulus!).
Second, there is a transition to the N-phase which is significantly softer than the
N one.

The reaction of the star structure to the (1st order) phase transition
(NH$\longrightarrow$N) can be described by the linear response theory
formulated  in \cite{ZdunikHS1987}. This theory describes stellar configurations in the vicinity
of the star with central pressure equal to $P_3$. In  our case this region of stellar
configurations is  very small since we are close to the maximum mass. The crucial
parameter, determining the stability  of the star with a  small core of the denser
 phase (N),  is the relative density jump at phase transition pressure, $P_3$:
$\lambda=\rho^\prime_3/\rho_3$. Stability condition for a star with a small N-core
reads:
\begin{equation}
\lambda<\lambda_{\rm crit}=\frac{3}{2}(1+\frac{P_3}{\rho_3 c^2})
\label{stabpt}
\end{equation}
(see Sect 3.4 of \citealt{ZdunikHS1987}). In our case condition (\ref{stabpt})
 is fulfilled, because $\lambda=1.27$ while $\lambda_{\rm crit}=2.18$. Consequently,
 there exists a  (very small) region of stable configurations with the
 N-phase core. Actually, this region is very narrow:
  the maximum mass of non-rotating stars is
reached for central pressure $P_{\rm c,max}$  larger only by 0.04\% than
the pressure at the phase transition, $P_3$.
\begin{figure}[h]
\resizebox{\columnwidth}{!} {\includegraphics[angle=-90]{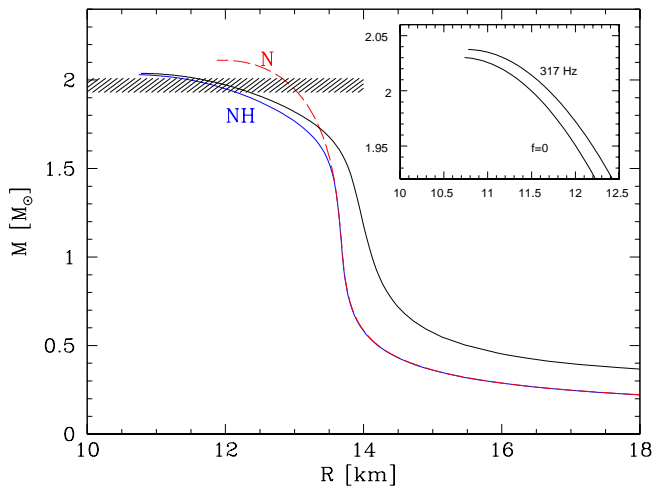}}
\caption{Gravitational stellar
mass, $M$,  versus circumferential radius, $R$, calculated for the EOS.N and
EOS.NH. Only stable configurations are displayed .
Inset: effect of rotation at $f=317$~Hz on the $M$ - equatorial
circumferential radius  curve near $M_{\rm max}$. }
\label{fig:mr_16}
\end{figure}

The $M-R$ relation for non-rotating NS, and those rotating
at 317 Hz,  based on the BM165 EOS,  are  plotted in
Fig.\;\ref{fig:mr_16}. Stars with $M>1.4~\msun$ have a hyperon core. The flattening
of the $M(R)$ curve due to the hyperon softening of the EOS is significant.
However, it still allows for $M_{\rm max}^{\rm (NH)stat}=2.03~\msun$, which is only
by $0.07~\msun$ smaller than $M_{\rm max}^{\rm (N)stat}$. Rotation at 317 Hz, as
measured for  PSR J1614-2230, increases $M_{\rm max}^{\rm (NH)}$ to $2.04~\msun$
(see zoomed inset  in Fig.\;\ref{fig:mr_16}).
\section{Discussion and conclusions}
\label{sect:conclusions}
We constructed a model of hyperon cores of neutron stars that allows for the
existence of neutron star of $2~\msun$. The model is consistent with ten
semi-empirical evaluations of nuclear and hyper-nuclear matter parameters.
As an additional constraint, we imposed SU(6) symmetry relations between the coupling
constants of baryons  and vector mesons. In spite of this, by introducing
two hidden-strangenes meson fields (scalar and vector) coupled to hyperons only,
we were able to reproduce four semi-empirical parameters stemming from hypernuclear
physics.

In contrast to NL3 model, used by \cite{Bonanno2011}, our symmetry energy is not
unusually "stiff" near the saturation point: we get $L=74$~MeV, compared to
$L=118~$MeV for NL3 (\citealt{Agrawal2005}). Consistently, our EOS.N is not
unusually stiff, and yields for NS with nucleon cores $M^{\rm(N)stat}_{\rm
max}=2.10~\msun$, to be contrasted with the NL3 value of $2.8~\msun$. The hyperon
softening for the model of \cite{Bonanno2011}) is dramatic, and leads to $M_{\rm max}^{\rm
(NH)stat}$ which is lower by nearly $0.8~\msun$ than the N one. In our case,
getting  $M_{\rm max}^{\rm (NH)stat}>2.0~\msun$ is conditioned by the high-density
vector interactions in the hyperon sector (not excluded in view of our lack of
knowledge of high-density hyperon interactions), while \cite{Bonanno2011} rely on
the assumed extreme stiffness of the  EOS.N, which is necessarily connected with a
very high value of their $L$.

Our EOS.NH becomes stiffer than EOS.N for $\rho\gtrsim 5\rho_0$, and its stiffness
grows with density. This leads eventually to the instability of the NH matter
against the conversion into the N one, softening the EOS due to the first order
phase transition. The maximum density at which stable NH phase can exist determines
actually our $M_{\rm max}^{\rm (NH)stat}$, which is only $0.07~\msun$ lower than
$M_{\rm max}^{\rm (N)stat}$. Rotation at $317~$Hz, measured for PSR J1614-2230,
increases $M_{\rm max}^{\rm (NH)}$ by $0.01~\msun$, to $2.04~\msun$.
Breaking the SU(6) symmetry for the vector-meson couplings to hyperons,
similarly as in the recent work of \cite{Debarati2011}, can make
 the value of $M_{\rm max}^{\rm (NH)}$ significantly higher
 (Bednarek et al., in preparation).

In the present paper we restricted to the hadronic forms of matter. A consistent
treatment of the phase transition to quark phase in neutron star core would require
the  use of the QCD for both hadronic and quark phases. As the transition  occurs in the
strong-coupling regime, approximations are not controllable. An approach based on
an effective model of the QCD of quark matter (Nambu-Jona-Lasinio) and NL3 for
hadronic phase, used by \cite{Bonanno2011} indicates that to get  $M_{\rm
max}>2~\msun$ vector repulsion in quark matter should be sufficiently strong. In
any case, the maximum mass obtained by them  is very close to that reached at
central density equal to the deconfinement density.

\acknowledgements
We are deeply grateful to Mikhail E. Gusakov for his precious help in
detecting and removing some errors in the earlier versions of the
present work. We also thank him for asking difficult but inspiring
questions referring to the physics of dense matter.

This work was partially supported by the Polish MNiSW
research grant no.N N203 512838, by the LEA Astro-PF, and
by the European Science Foundation CompStar RNP.  MB acknowledges the support
of Marie Curie Fellowship no. ERG-2007-224793 within the 7th European
Community Framework Programme.



\begin{thebibliography}{}{

\bibitem[Agrawal et al.(2005)]{Agrawal2005}
Agrawal, B.K., Shlomo, S., Kim Au, V., 2005, Eur Phys J A, 25, 525


\bibitem[Akmal et al.(1998)]{AkmalPR1998}
Akmal, A., Pandharipande, V.~R., Ravenhall, D.~G.,
 1998, Phys Rev C, 58, 1804


\bibitem[Baym et al. (1971)]{BPS1971}
Baym, G., Pethick, C., Sutherland, P., 1971, ApJ, 170, 299

\bibitem[Bednarek \& Ma\'nka(2009)]{BednarekM2009}
Bednarek, I., \& Manka, R.\ 2009, Journal of Physics G: Nuclear Physics,
36, 095201

\bibitem[Bonanno \& Sedrakian (2011)]{Bonanno2011}
Bonanno, L., Sedrakian, A., 2011,
arXiv:1108.0559v1 [astro-ph.SR] 2 Aug 2011


\bibitem[Bonazzola et al.(1993)]{BGSM1993}
Bonazzola S., Gourgoulhon E., Salgado M., Marck J.-A., 1993,
A \& A, 278, 421



\bibitem[Bombaci et al. (2008)]{Bombaci2008}
Bombaci, I., Panda, P.K., Providencia, C., Vidana, I., 2008,
 Phys Rev D, 77, 083002



\bibitem[Burgio et al. (2011)]{Burgio2011}
Burgio, G.F., Schulze, H.-J., Li, A., 2011,
Phys Rev C, 83, 025804


\bibitem[Chatterjee \& Schaeffner-Bielich (2011)]{Debarati2011}
Chatterjee D., Schaeffner-Bielich J., 2011, poster presented at
CompStar2011 Workshop, Catania, Italy


\bibitem[Demorest et al. (2010)]{Demorest2010}
Demorest P.B., Pennucci T., Ransom S.M., Roberts M.S.E., Hessels J.W.T.,
2010, Nature, 467, 1081

\bibitem[Douchin \& Haensel(2001)]{DouchinH2001}
Douchin, F., Haensel, P., 2001, A\& A, 380, 151

\bibitem[Endo et al. (2006)]{Endo2006}
Endo, T., Maruyama, T., Chiba, S., Tatsumi, T.,
2006, Prog Theor Phys, 115, 337


\bibitem[Glendenning (1985)]{Glend1985}
Glendenning, N.K., 1985,
 ApJ, 293, 470


\bibitem[Glendenning \& Moszkowski (1991)]{GlendMoszk1991}
Glendenning, N.K., Moszkowski, S.A., 1991,
 Phys Rev Lett, 67, 2414

\bibitem[Glendenning (1996)]{GlendBook}
Glendenning, N.K., 1996, Compact Stars. Nuclear Physics, Particle
Physics, and General Relativity  (Springer, New York)


\bibitem[Haensel \& Pichon (1994)]{HaenselP1994}
Haensel, P., Pichon, B., 1994, A\& A, 283, 313


\bibitem[Haensel et al.(2007)]{NSbook2007}
Haensel, P., Potekhin, A.Y., Yakovlev, D.G. 2007,
Neutron Stars 1. Equation of State and Structure (New York, Springer)

\bibitem[Horowitz \& Piekarewicz (2001)]{HorowitzPiekarewicz2001}
Horowitz, C.J., Piekarewicz, J., 2001, Phys Rev Lett, 86, 5647


\bibitem[Rikovska Stone et al. (2007)]{RikovskaStone2007}
Rikovska Stone, J., Guichon, P.A.M., Matevosyan, H.H., Thomas, A.W., 2007,
Nucl. Phys. A, 792, 341

\bibitem[Schaffner et al. (1994)]{SchaffnerDover1994}
Schaffner, J., Dover, C.B., Gal, A., Millener, D.J., Greiner, C. Stocker, H.,
 1994, Ann Phys (NY), 235, 35


\bibitem[Schaffner-Bielich \& Gal (2000)]{SchaffnerGal2000}
Schaffner-Bielich, J., Gal, A., 2000, Phys Rev C, 62, 034311

\bibitem[Song et al. (2003)]{SongSu2003}
Song, H.Q., Su, R.K., Lu, D.H., Qian, W.L., 2003, Phys Rev C, 68, 055201


\bibitem[Stone et al. (2010)]{Stone2010}
Stone, J., Guichon, P.A.M., Thomas, A.W., 2010,
arXiv:1012.2919v1 [nucl-th] 14 Dec
2010

\bibitem[Sugahara \& Toki (1994)]{SugaharaToki1994}
Sugahara, Y., Toki, H., 1994, Nucl Phys A, 579, 557

\bibitem[Takahashi et al. (2001)]{Takahashi2001}
Takahashi, H., et al., 2001, Phys Rev Lett, 87, 212502

\bibitem[Vidana et al. (2011)]{Vidana2011}
Vidana, I., Logoteta, D., Providencia, C., Bombaci, I., 2011,
Europhys. Lett., 94, 11002

\bibitem[Walecka (2004)]{Walecka2004}
Walecka J.D., 2004, Theoretical Nuclear and Subnuclear Physics,
second edition (World Scientific - Imperial College Press)

\bibitem[Zdunik et al.(1987)]{ZdunikHS1987}
Zdunik, J.~L., Haensel, P., \& Schaeffer, R., 1987, A\& A, 172, 95 }
\end{thebibliography}
\end{document}